\documentclass[11pt]{article}

\RequirePackage[OT1]{fontenc}
\RequirePackage{amsthm,amsmath}
\RequirePackage[numbers]{natbib}
\usepackage[dvipsnames]{xcolor}
\usepackage{hyperref}
\usepackage{graphicx}
\usepackage{amssymb}
\usepackage{fullpage}

\newcommand{\argmin}{\operatorname*{arg \ min}}

\newtheorem{remark}{Remark}

\begin{document}

\title{A covariance-enhanced approach to multi-tissue joint {e}QTL mapping with application to transcriptome-wide association studies}

\author{Aaron J. Molstad$^1$\footnote{amolstad@ufl.edu}, Wei Sun$^{2,3,4}$, Li Hsu$^{2,3}$\\
University of Florida$^1$, Fred Hutchinson Cancer Research Center$^2$, \\University of Washington$^3$, University of North Carolina-Chapel Hill$^4$}
\date{}
\maketitle




\begin{abstract}
Transcriptome-wide association studies based on genetically predicted gene expression have the potential to identify novel regions associated with various complex traits. It has been shown that incorporating expression quantitative trait loci (eQTLs) corresponding to multiple tissue types can improve power for association studies involving complex etiology. In this article, we propose a new multivariate response linear regression model and method for predicting gene expression in multiple tissues simultaneously. Unlike existing methods for multi-tissue joint eQTL mapping, our approach incorporates tissue-tissue expression correlation, which allows us to more efficiently handle missing expression measurements and more accurately predict gene expression using a weighted summation of eQTL genotypes. We show through simulation studies that our approach performs better than the existing methods in many scenarios. We use our method to estimate eQTL weights for 29 tissues collected by GTEx, and show that our approach significantly improves expression prediction accuracy compared to competitors. Using our eQTL weights, we perform a multi-tissue-based S-MultiXcan \citep{barbeira2019integrating} transcriptome-wide association study and show that our method leads to more discoveries in novel regions and more discoveries overall than the existing methods. Estimated eQTL weights are available for download online at \texttt{github.com/ajmolstad/MTeQTLResults}
\end{abstract}



\section{Introduction}
Genome-wide association studies (GWAS) have identified tens of thousands of reproducible trait associated single-nucleotide polymorphisms (SNPs) through agnostic SNP-by-SNP association analysis (see \href{https://www.ebi.ac.uk/gwas/}{https://ebi.ac.uk/gwas/}) \cite{buniello2019nhgri}.  Though most of these associated SNPs lie outside of any gene, they are enriched for expression quantitative trait loci (eQTL)\citep{hindorff2009potential, maurano2012systematic, nicolae2010trait}, which are genetic loci that affect the expression of one or more genes \citep{cheung2003natural, montgomery2010transcriptome, pickrell2010understanding, schadt2003genetics, stranger2005genome}. Machine learning methods have been used to infer eQTL regulation of a gene using all nearby genetic variants \citep{gamazon2015gene, manor2015genoexp, zeng2017prediction}. Using genetically predicted gene expression from these models, researchers have performed transcriptome-wide association studies (TWAS) and reported many novel regions associated with various complex traits \citep{gusev2016,pavlides2016predicting, mancuso2017integrating, barbeira2018exploring}, many of which have no GWAS association within 1Mb. 
There are several advantages to such analyses: leveraging gene expression enriches potential trait associated SNPs, aggregating signals through joint eQTL modeling enhances the overall association strength, and the number of tests is substantially reduced from testing millions of SNPs to about 20,000 genes. Because of the tissue-dependent nature of gene expression, these analyses typically use gene expression from a single trait-relevant tissue. However, recent works have shown that eQTLs are often shared across multiple tissues and assessing the association of genetically predicted gene expression using multiple tissues improves power for genetic association with complex traits \citep{barbeira2018exploring, barbeira2019integrating}.

Leveraging shared eQTLs across tissues improves power for eQTL discoveries and gene expression imputation accuracy, which can, in turn, further improve power for transcriptome-wide association analysis. \citet{flutre2013statistical} and \citet{li2017empirical} proposed  multi-tissue eQTL mapping approaches that identified more eQTLs than tissue-by-tissue approaches. More recently, \citet{hu2018statistical} proposed a penalized regression approach for joint modeling of eQTLs using a penalty which encourages shared eQTLs across tissues. Using the genotype and expression data for various tissues from the Genotype-Tissue Expression (GTEx) project \citep{gtex2015genotype, gtex2017genetic}, they showed multi-tissue eQTL models improve imputation accuracy and gene association detection substantially compared to single-tissue approaches. 

\begin{figure}[t!]
\begin{center}
\makebox[\textwidth][c]{\includegraphics[width=1.1\textwidth]{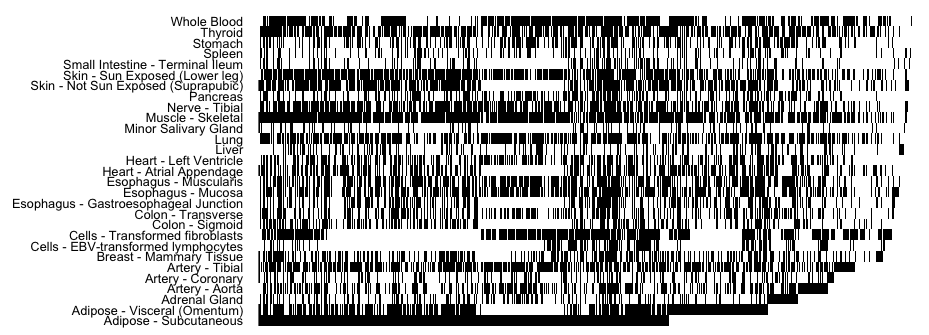}}
\end{center}
\caption{
A heatmap of missingness in GTEx gene expression data for the 29 tissues we analyzed. Rows correspond to the 29 tissues and columns correspond to the 613 subjects with expression measured in at least one of the corresponding tissues. White spaces denote missing measured expression, whereas black denote observed measured expression.}\label{fig:missingnessGTEx}
\end{figure}

However, the method of \citet{hu2018statistical} does not take into account tissue-tissue correlation of gene expression in joint eQTL modeling. In the recent statistical literature, it has been shown that in high-dimensional penalized multivariate response linear regression, accounting for error correlation (here, tissue-tissue correlation) often leads to improved variable selection and prediction accuracy \citep{rothman2010sparse,yin2011sparse,wang2015joint,lee2012simultaneous}. This phenomenon can be partly explained by a seemingly unrelated regression interpretation of high-dimensional sparse multivariate response linear regression \citep{zellner1962efficient, srivastava1987seemingly}. Further, owing to biological and cost constraints, some tissue types are harder to obtain than others. For example, of the 29 GTEx tissues we focus on in our analysis, there were 613 individuals with expression measured in at least one tissue of interest. Some individuals have as few as one tissue with measured expression, whereas others have expression measured in as many as 28 tissues. No individual has expression measured in all 29 tissues. The sample size per tissue varies from 85 (Minor Salivary Gland) to 491 (Muscle - Skeletal) (Figure~\ref{fig:missingnessGTEx}). We show that leveraging tissue-tissue correlation can substantially improve gene expression prediction accuracy, especially for tissues with small sample sizes. We do so by imputing the missing gene expression in a way which simultaneously estimates tissue-tissue correlation and joint eQTL weights: the mechanism through which this operates is straightforward, for example, if two tissues' expression is highly correlated, measuring expression in only one of these tissues allows one to reasonably estimate expression in the unmeasured tissue. If tissue-tissue correlation is ignored, substantial gains in gene expression prediction accuracy may be lost. 

In this article, we propose a new method for multi-tissue joint eQTL mapping which can be used when individuals have missing gene expression measurements in many tissues. We develop an efficient penalized expectation-conditional-maximization (ECM) algorithm to solve the optimization problem. Our penalties allow us to identify both tissue specific and shared eQTLs while simultaneously modeling cross-tissue expression covariance. Compared to existing methods for multi-tissue joint eQTL mapping, our approach has several advantages: 
\begin{itemize} 
  \item our method explicitly models tissue-tissue correlation, providing new insights about cross-tissue expression dependence which cannot be explained by eQTLs; 
  \item by modeling cross-tissue correlation, our method more efficiently makes use of the available expression measurements for eQTL weight estimation;
   \item our computational approach can be used to more efficiently compute special cases of our method, e.g., the method of \citet{hu2018statistical};
  \item in both simulations and our analysis of the GTEx data, our approach leads to improved gene expression prediction accuracy compared to (a) tissue-by-tissue approaches which estimate eQTL weights separately; (b) two-step approaches where imputation and prediction are performed in sequential steps; and (c) approaches which ignore cross-tissue correlation.
\end{itemize}
The implication of the final point on association analyses of genetically predicted gene expression is immediate: with more accurate expression prediction models, one can perform more reliable tests, and thus, can expect more novel regions to be discovered. In particular, with the weights estimated from the GTEx data using our method, we performed a multi-tissue-based S-MultiXcan\citep{barbeira2019integrating} TWAS analysis of four complex traits. We found that our weights led to more total discoveries and more novel discoveries beyond single variant analyses than existing methods in four traits we studied. Focusing on genes associated with the occurrence of a heart attack, we identified multiple regions which were not attributable to any GWAS associated SNP, but have been associated with traits related to heart function, e.g., coronary heart disease, triglyceride levels, and LDL cholesterol. These findings demonstrate the potential power gain using an integrative analysis with multiple tissues and also offer insight into potential factors associated with the occurrence of a heart attack. Discovered genes from our S-MultiXcan analysis, along with our estimated eQTL weights, are available for download at \texttt{github.com/ajmolstad/MTeQTLResults}.

\section{Method}
\subsection{Penalized maximum likelihood estimator}
Throughout, we will assume a general model for cross-tissue expression given SNP genotypes. For a particular gene, let $y_i \in \mathbb{R}^q$ denote the vector of centered and normalized measured expression in $q$ tissues for the $i$th subject and let $x_i \in \mathbb{R}^p$ denote the $p$ SNP genotypes (also centered and normalized) within a certain distance (e.g., 500kb) of the transcription start and end site of the gene of interest. We assume that for the $i$th subject, measured expression is a realization of the random vector
\begin{equation}\label{eq:Normal_model}
\mathcal{Y}_i = \beta_*'x_i + \epsilon_i, \quad\quad \epsilon_i \sim {\rm N}_q\left(0, \Omega_*^{-1}\right), \quad\quad (i = 1, \dots, n) 
\end{equation}
where ${\rm N}_q$ denote the $q$-dimensional multivariate normal distribution,  $\beta_* \in \mathbb{R}^{p \times q}$ is the unknown regression coefficient matrix (i.e., eQTL weights), and $\Omega_* \in \mathbb{S}^{q}_+$ is the cross-tissue error precision (inverse covariance) matrix. We further assume that $\epsilon_i$ is independent of $\epsilon_j$ for all $i \neq j$. Throughout, we let $\mathbb{S}^{q}_+$ denote the set of $q \times q$ symmetric and positive definite matrices. 

We estimate $\beta_*$ and $\Omega_*$ jointly by minimizing the observed-data penalized negative log-likelihood with respect to $\beta$ and $\Omega$, the optimization variables corresponding to $\beta_*$ and $\Omega_*$, respectively. Let $o_i$ and $m_i$ denote the components of $y_i$ (tissues) which are observed and missing, respectively, so that without loss of generality, we can write $y_i = (y_{i,o_i}', y_{i,m_i}')' \in \mathbb{R}^q$ where $A'$ denotes the transpose of matrix or vector $A$ and $y_{i,o_i}$ denotes the subvector of $y_i$ corresponding to the indices $o_i$. 
Thus, the observed-data negative log-likelihood (ignoring constants) for the $n$ subjects is proportional to
\begin{equation} \label{eq:negloglik}
\frac{1}{n}\sum_{i=1}^n \left[ {\rm tr}\left\{(y_{i,o_i} - x_i'\beta_{\cdot,o_i})' \Sigma_{o_i}^{-1}(y_{i,o_i} - x_i'\beta_{\cdot,o_i}) \right\} + \log {\rm det}\left(\Sigma_{o_i}\right) \right],
\end{equation}
where $\Sigma_{o_i} \in \mathbb{S}^{\# o_i}_+$ is the submatrix of $\Sigma$ corresponding to the indices $o_i$, $\beta_{\cdot, o_i} \in \mathbb{R}^{p \times {\#}o_i}$ is the submatrix of $\beta$ containing columns corresponding to the indices $o_i$, and $\# o_i$ denotes the cardinality of $o_i$. 
Throughout, ${\rm tr}$ and ${\rm det}$ denote the trace and determinant operators, respectively. Note that for notational simplicity, we assume that $y_i$ and $x_i$ have column-wise average zero, so that we can write \eqref{eq:Normal_model} without an intercept.

Unfortunately, minimizing \eqref{eq:negloglik} directly is computationally difficult since missingness patterns differ across subjects. Instead, we use an expectation-maximization (EM) algorithm, which allows us to operate on the complete-data negative log-likelihood. Let $Y_O$ denote the collection of the $y_{i,o_i}$'s for $i=1, \dots, n$, i.e., $Y_O$ is the collection of all observed gene expression for the $n$ subjects.  Similarly, let $Y_M$ denote the collection of all unmeasured (missing) gene expression for the $n$ subjects. Thus, we rewrite the observed-data negative log-likelihood $\mathcal{F}(\beta, \Omega; Y_O)$ in terms of the complete-data negative log-likelihood:
\begin{equation}\label{log_likelihood} \mathcal{F}(\beta, \Omega; Y_O) =  - \frac{1}{n}\sum_{i=1}^n  \log  \int f_{\mathcal{Y}}(\beta, \Omega; y_{i,o_i}, y_{i,m_i}) \partial_{\mathcal{Y}_{i,m_i}} ,
\end{equation}
where $f_{\mathcal{Y}}$ denotes the $q$-dimensional multivariate normal density. 

To estimate $\beta_*$ and $\Omega_*$ simultaneously while accounting for missingness, we propose to minimize a penalized version of $\mathcal{F}(\beta, \Omega; Y_O)$ with respect to $\beta$ and $\Omega$ jointly. Specifically, we estimate $(\beta_*, \Omega_*)$ with
\begin{equation}\label{estimator}
\argmin_{\beta \in \mathbb{R}^{p \times q}, \Omega \in \mathbb{S}^q_{+}} \left\{ \mathcal{F}(\beta, \Omega; Y_O) + \lambda_\beta \mathcal{P}^{(\alpha)}_\beta(\beta) + \lambda_\Omega  \mathcal{P}_\Omega(\Omega) \right\},
\end{equation}
where $\lambda_\beta$ and $\lambda_\Omega$ are non-negative tuning parameters; and $\mathcal{P}^{(\alpha)}_\beta$ and $\mathcal{P}_\Omega$ are convex penalty functions applied to $\beta$ and $\Omega$, respectively. The penalties are chosen based on the biology underlying multi-tissue joint eQTL mapping. In particular, it is believed that large proportion of eQTLs are shared across multiple tissues in most genes \citep{flutre2013statistical,li2017empirical,hu2018statistical}, which would imply that nonzero entries of $\beta_*$ are likely to occur in a subset of rows since each row of $\beta_*$ corresponds to a particular SNP's regression coefficients (eQTL weights) for the $q$ tissues. To exploit this assumption, we use a penalty which balances row-wise sparsity with element-wise sparsity:
\begin{equation}\label{eq:beta_pen}
 \mathcal{P}^{(\alpha)}_\beta(\beta) = \sum_{j=1}^p \left\{ \alpha \left(\sum_{k=1}^q|\beta_{j,k}|\right) + (1-\alpha) \|\beta_{j,\cdot}\|_2\right\},
 \end{equation}
where $\alpha \in [0,1]$ is a tuning parameter, $\|\cdot\|_2$ denotes the Euclidean norm of a vector, and $\beta_{j, \cdot} \in \mathbb{R}^q$ is the $j$th row of $\beta$ $(j = 1, \dots, p)$. If $\alpha$ were set to zero, estimates of $\beta_*$ would only have rows which are entirely zero or nonzero. Conversely, when $\alpha=1$, this penalty does not encourage eQTL sharing explicitly. The penalty in \eqref{eq:beta_pen} was also used by \citet{peng2010regularized} in the context of the multivariate response linear regression model of gene expression on copy number alterations. 

To estimate the cross-tissue error precision matrix $\Omega_*$, we use an $\ell_1$-penalty on the entries of the corresponding optimization variable:
$\mathcal{P}_\Omega(\Omega) = \sum_{j,k} |\omega_{j,k}|.$
For sufficiently large values of the tuning parameter $\lambda_\Omega$, this penalty leads to estimates of the precision matrix which will have all off-diagonal entries equal to zero \citep{yuan2007model,rothman2008sparse}. Hence, this penalty implicitly assumes that some entries of $\Omega_*$ equal zero. This assumption is also biologically reasonable in the context of multi-tissue joint eQTL mapping: it is well known that under  \eqref{eq:Normal_model}, a zero in the $(j,k)$th entry of $\Omega_*$ implies that expression in the $j$th and $k$th tissues are independent given expression in all other tissues and all $p$ SNP genotypes.

\subsection{Penalized expectation-conditional-maximization algorithm}
To obtain the penalized maximum likelihood estimator, i.e., to solve \eqref{estimator} with the penalties in \eqref{eq:beta_pen} and $\mathcal{P}_\Omega(\Omega)$, we propose a penalized expectation-conditional-maximization (ECM) algorithm \citep{meng1993maximum}. Let $\mathcal{G}_\lambda$ denote the penalized complete-data negative log-likelihood so that
$$\mathcal{G}_\lambda(Y_O, Y_M; \beta, \Omega) = - \frac{1}{n} \sum_{i=1}^n \log  f_\mathcal{Y}(\beta, \Omega; y_{i, o_i}, y_{i, m_i}) + \lambda_\beta \mathcal{P}_\beta^{(\alpha)}(\beta) + \lambda_\Omega \mathcal{P}_{\Omega}(\Omega)$$
where  $\lambda = (\alpha, \lambda_\beta, \lambda_\Omega) \in [0,1] \times \mathbb{R}_+ \times \mathbb{R}_+$ is fixed. The standard EM algorithm proceeds in two steps: the E-step
$$ Q(\beta, \Omega \mid \beta^{(k)}, \Omega^{(k)}) = \mathbb{E}\left[ \mathcal{G}_\lambda(Y_O, Y_M; \beta, \Omega) \mid Y_O, \beta^{(k)}, \Omega^{(k)} \right],
$$
and the subsequent M-step
\begin{equation}\label{M_step_1} 
(\beta^{(k+1)}, \Omega^{(k+1)}) \in \hspace{-2pt} \argmin_{\beta \in \mathbb{R}^{p \times q}, \Omega \in \mathbb{S}_+^q}\hspace{-2pt} Q(\beta, \Omega \mid \beta^{(k)}, \Omega^{(k)}). 
\end{equation}
However, solving \eqref{M_step_1} exactly requires a blockwise coordinate descent algorithm iterating between updating $\Omega$ and $\beta$. Instead, we propose to update each variable once (with the other held fixed) and then return to the E-step. In this way, our approach can be considered a \textit{generalized} expectation-conditional-maximization algorithm in the sense that we do not solve the M-step exactly at each iteration, but are guaranteed that the objective function is non-increasing. An outline of the complete algorithm is given in Algorithm 1. \\

\noindent \textit{Algorithm 1: (Penalized ECM)} Initialize optimization variables. Set $k=1$. 
\begin{enumerate}
  \item Compute $Q(\beta, \Omega \mid \beta^{(k)}, \Omega^{(k)}) =   \mathbb{E} \left[ \mathcal{G}_\lambda(Y_O, Y_M; \beta, \Omega) \mid Y_O, \beta^{(k)}, \Omega^{(k)} \right]$
  \item Compute $\Omega^{(k+1)} \leftarrow \argmin_{\Omega \in \mathbb{S}^q_+} Q(\beta^{(k)}, \Omega \mid \beta^{(k)}, \Omega^{(k)}).$
  \item Compute $\beta^{(k+1)} \leftarrow \argmin_{\beta \in \mathbb{R}^{p \times q}} Q(\beta, \Omega^{(k+1)} \mid \beta^{(k)}, \Omega^{(k)}).$
  \item If the objective value from \eqref{estimator} has not converged, set $k = k+1$ and return to Step 1.
\end{enumerate}
\bigskip

Because the updates for $\Omega$ and $\beta$ (with the other held fixed) are both convex optimization problems, the M-step is an instance of a \textit{biconvex} optimization problem.
In the following subsection, we describe how to solve Steps 1-3 of the penalized ECM algorithm.

\subsection{Algorithm details}
To compute the $Q$ function in Step 1 (E-Step) of Algorithm 1, we use the conditional multivariate normal distribution for $y_{i,m_i}$ given $y_{i,o_i}$ and the current iterates of $\beta$ and $\Omega$:
\begin{equation}\label{eq:ConditionalNormal}
(y_{i,m_i} \mid y_{i,o_i}, \beta^{(k)}, \Omega^{(k)})
\sim {\rm N}_{\# m_i}\left(\mu_{i, m_i}^{(k)}, V_{i, m_i}^{(k)}\right),\quad i=1,\dots, n,
\end{equation}
where the mean and variance from \eqref{eq:ConditionalNormal} can be expressed in closed form:
\begin{align*}
  \mu_{i, m_i}^{(k)} &= {\beta^{(k)}_{\cdot, m_i}}'x_i + \Sigma_{m_i, o_i}^{(k)}\Sigma_{o_i}^{(k)-1}(y_{i,o_i} - {\beta^{(k)}_{\cdot, o_i}}'x_i);\\
  V_{i, m_i}^{(k)} &= \Sigma_{m_i}^{(k)} - \Sigma_{m_i, o_i}^{(k)} \Sigma_{o_i}^{(k)-1}\Sigma_{o_i, m_i}^{(k)}
\end{align*}
with $\Sigma_{m_i, o_i}^{(k)}$ denoting the submatrix of $\Sigma^{(k)}$ whose rows correspond to the indices of $m_i$ and whose columns correspond to indices of $o_i$, i.e., $\Sigma^{(k)}_{o_i} \equiv \Sigma^{(k)}_{o_i,o_i}$ and  $\Sigma^{(k)}_{m_i} \equiv \Sigma^{(k)}_{m_i,m_i}$.  

With $\mu_{i, m_i}^{(k)}$ and $V_{i, m_i}^{(k)}$ computed and stored for $i=1, \dots, n$, we can then express Step 2 of Algorithm 1 as a familiar convex optimization problem. 
\begin{remark} Let $\beta^{(k)}$ and $\Sigma^{(k)}$ be fixed. Then, Step 2 of Algorithm 1 can be expressed as
\begin{equation}\label{eq:glasso}
\Omega^{(k+1)} = \argmin_{\Omega \in \mathbb{S}_+^p} \left[ {\rm tr}\left\{S(\beta^{(k)}, \Sigma^{(k)})\Omega\right\}- \log {\rm det}(\Omega) + \lambda_\Omega \sum_{j,k}|\omega_{j,k}| \right] \end{equation}
where $S(\beta^{(k)}, \Sigma^{(k)}) = n^{-1} \sum_{i=1}^n \Gamma_i^{(k)}$ for $i=1, \dots, n$ with the submatrices of each $\Gamma_i^{(k)}$ equal to
   \begin{align*}
[\Gamma_{i}^{(k)}]_{o_i} &= (y_{i,o_i}' - x_i'\beta^{(k)}_{\cdot, o_i})'(y_{i,o_i}' - x_i'\beta^{(k)}_{\cdot,o_i});\\
[\Gamma_{i}^{(k)}]_{m_i} &= (\mu_{i, m_i}^{(k)'} -   x_i'\beta^{(k)}_{\cdot,m_i})'(\mu_{i, m_i}^{(k)'} -   x_i'\beta^{(k)}_{\cdot,m_i}) + V_{i, m_i}^{(k)};\\
[\Gamma_{i}^{(k)}]_{o_i, m_i} &= (y_{i,o_i}' - x_i'\beta^{(k)}_{\cdot, o_i})'(\mu_{i, m_i}^{(k)} -   x_i'\beta^{(k)}_{\cdot,m_i}). 
\end{align*}
\end{remark}
Conveniently, \eqref{eq:glasso} is exactly the optimization problem for computing the $\ell_1$-penalized normal log-likelihood precision matrix estimator with input sample covariance matrix $S(\beta^{(k)}, \Sigma^{(k)})$. Many efficient algorithms and software packages exist for computing \eqref{eq:glasso}: in our implementation, we use \texttt{glasso} in \texttt{R} \citep{witten2011new}.  

Step 3 of Algorithm 1, the update for $\beta$ with $\Omega^{(k+1)}$ fixed, can be expressed as a minimizer of a penalized weighted residual sum of squares criterion, which we formalize in Remark 2.
\begin{remark} Let $\beta^{(k)}$, $\Omega^{(k)}$, and $\Omega^{(k+1)}$ be fixed. Then, Step 3 of Algorithm 1 can be expressed as
\begin{equation} \label{beta_update}
\beta^{(k+1)} = \argmin_{\beta \in \mathbb{R}^{p \times q}}  \left[\frac{1}{n} {\rm tr}\left\{(\bar{Y}^{(k)} - X\beta) \Omega^{(k+1)}(\bar{Y}^{(k)} - X\beta)'\right\} + \lambda_\beta \mathcal{P}^{(\alpha)}_\beta(\beta) \right]
\end{equation}
where $\bar{Y}^{(k)} \in \mathbb{R}^{n \times q}$ has $i$th row $\bar{Y}^{(k)}_{i,(o_i, m_i)} = (y_{i,o_i}', {\mu_{i,m_i}^{(k)'}})'$ for $i=1, \dots, n$.
\end{remark}
Note that the notation $Y_{i,(a)}$ denotes the permutation of the elements of the vector $Y_{i}$ corresponding to an index set $a$. For example, were $a = (2,1)$ and $U_i \in \mathbb{R}^2$, $U_{i,(a)} = (U_{i,2}, U_{i,1})'$.  

To solve \eqref{beta_update}, we use an accelerated proximal gradient descent algorithm \citep{parikh2014proximal}. We briefly motivate this iterative procedure from a majorize-minimize perspective \citep{lange2016mm}. Let $h:\mathbb{R}^{p \times q} \to \mathbb{R}$ denote the unpenalized objective function from \eqref{beta_update}, i.e., the objective function from \eqref{beta_update} with $\lambda_\beta = 0$.  Let $\|A\|_F^2 = {\rm tr}(A'A) = \sum_{j,k} A_{j,k}^2$ denote the squared Frobenius norm of a matrix $A$. Then, given some iterate of $\beta$ which is fixed, say $\beta^{(r)}$, because $h$ is convex and has Lipschitz continuous gradient, 
$$ h(\beta)  \leq h(\beta^{(r)}) + {\rm tr}\left\{\nabla h(\beta^{(r)})'(\beta - \beta^{(r)})\right\} + \frac{1}{2\gamma}\|\beta - \beta^{(r)}\|_F^2$$
for all $\beta$ with $\gamma > 0$ sufficiently small. Hence, it follows that
\begin{align}\label{eq:MM} 
h(\beta) + \lambda_\beta \mathcal{P}^{(\alpha)}_\beta(\beta) \leq h(\beta^{(r)}) + {\rm tr}& \left\{ (\beta - \beta^{(r)})'\nabla h(\beta^{(r)}) \right\} + \frac{1}{2\gamma} \|\beta^{(r)} - \beta\|_F^2 + \lambda_\beta \mathcal{P}^{(\alpha)}_\beta(\beta)
\end{align}
for all $\beta$ where $\nabla h(\beta^{(r)})$ is the gradient of $h$ evaluated at $\beta^{(r)}$. Thus, if we minimize the right hand side of \eqref{eq:MM} with respect to $\beta$, which after some algebra can be expressed as
 \begin{equation}\label{eq:prox_update}
 \beta^{(r+1)} = \argmin_{\beta \in \mathbb{R}^{p \times q}} \left\{ \frac{1}{2}\| \beta - \beta^{(r)} + \gamma \nabla h(\beta^{(r)})\|_F^2 + \gamma \lambda_\beta \mathcal{P}^{(\alpha)}_\beta(\beta) \right\},
 \end{equation}
 we are guaranteed that the objective function from \eqref{beta_update} evaluated at $\beta^{(r+1)}$ is less than or equal to the objective function evaluated at $\beta^{(r)}$. This suggests a simple iterative procedure to solve \eqref{beta_update}: in the first step, we construct the ``majorizing function''\citep{lange2016mm} from the right hand side of \eqref{eq:MM} at the current iterate; in the second step, we minimize this majorizing function to obtain our new iterate; and then we repeat these two steps until the objective function from \eqref{beta_update} converges. 

This approach is computationally efficient because \eqref{eq:prox_update} can be solved in closed form with the following steps:
\begin{enumerate}
\item[(a)] Compute $\Delta = \beta^{(r)} - \gamma \nabla h(\beta^{(r)})$; 
\item[(b)] Compute $\bar{\Delta}_{j,k} = {\rm max}\left(|\Delta_{j,k}| - \gamma \lambda_\beta  \alpha, 0\right){\rm sign}(\Delta_{j,k})$ for all $(j,k) \in [1, \dots, p] \times [1, \dots, q]$
\item[(c)] Compute $\beta^{(r+1)}_{j,\cdot} = {\rm max}\left(1 - \frac{ \gamma \lambda_\beta (1-\alpha)}{\|\bar{\Delta}_{j,\cdot}\|_2}, 0\right)\bar{\Delta}_{j, \cdot}$ for $j = 1, \dots, p$ 
\end{enumerate}
For the complete algorithm to solve \eqref{beta_update}, steps (a) - (c) are repeated until the objective function converges. In the Supplementary Material, we give the exact steps of the accelerated version this algorithm which we use in our implementation and discuss selecting the step size $\gamma$. 

\subsection{Related methods}
A special case of our method was proposed by \citet{hu2018statistical}. In particular, the objective function they propose to estimate eQTL weights is equivalent to \eqref{estimator} with the constraint that $\Omega = I_q$, i.e., they implicitly assume that gene expression is uncorrelated with equal variance after conditioning on SNP genotypes. 
However, \citet{hu2018statistical} do not solve the optimization problem they posed directly. Instead, they devised an efficient coordinate descent scheme which approximated their estimator. While their approximation performed well in terms of predicting expression, 
comparing our method to theirs directly is difficult: it is not clear when to terminate their iterative procedure because their algorithm does not minimize an objective function whose value can be computed. Hence, we compare our approach, which assumes only $\Omega \in \mathbb{S}^q_+$, to what we refer to as the exact version of the \citet{hu2018statistical} 
(i.e., \eqref{estimator} with the constraint $\Omega = I_q$) in our simulation studies, the GTEx data analysis, and the S-MultiXcan TWAS. To compute the estimator of \citet{hu2018statistical}, we use a proximal gradient descent algorithm similar to that used for the optimization problem in \eqref{beta_update}. Details about this algorithm are provided in the Supplementary Material.

\section{Simulation studies}
\subsection{Data generating models and competing methods}
We performed extensive numerical experiments to study how the number of shared eQTLs, the population $R^2$, and tissue-tissue correlation structure affect the performance of various methods for estimating multi-tissue eQTL weights. 

To closely mimic the settings of the joint eQTL mapping in the GTEx data, we obtained whole genome sequencing SNP genotype data for all SNPs within 500kb of the BRCA1 gene for 620 subjects from the GTEx dataset. After pruning SNPs with extremely high correlations (see Data Preparation section), we are left with $p = 1178$ SNP genotypes. For each replication, we then generated $n = 620$ subjects' expression in $q=29$ tissues: letting $x_i \in \mathbb{R}^{p}$ be the SNP genotypes for the $i$th subject, we generated $y_i \in \mathbb{R}^q$, as a realization of the random vector 
$$ \beta_*'x_i + \epsilon_i, 
\quad i = 1, \dots, n,$$
where $\beta_* \in \mathbb{R}^{p \times q}$ are the eQTL weights and $\epsilon_i \sim {\rm N}_q(0, \Omega_*^{-1})$ are independent and identically distributed errors. For one hundred independent replications in each setting, we randomly split the $n = 620$ subjects into a training set of size $n_{\rm train} = 400$, a validation set of size 110, and a testing set of size 110. 

Independently in each replication, we generated $\beta_*$ as follows: first, we generated $B \in \mathbb{R}^{p \times q}$ to have entries which were independent ${\rm N}(0, 1)$.  Then, we generated $S \in \mathbb{R}^{p \times q}$ to be a matrix whose rows are either all zero or all one: we randomly select $s$ rows to be nonzero, where $s \in \left\{1,\dots, 20\right\}$. Given $S$, we then generated $U \in \mathbb{R}^{p \times q}$ so that each of the $q$ columns has 20-$s$ randomly selected entries equal to one only from entries which are zero in $S$ and all others equal to zero. With these, we set $\beta_* = B \circ S + B \circ U$ where $\circ$ is the elementwise product. By constructing $\beta_*$ in this way, each tissue has twenty total eQTLs, $s$ of which are shared across all tissue types. We consider $s = \left\{5, 10, 15, 18, 20\right\}$ in the simulations we present in this section.  Note that since many SNPs are highly correlated (linkage disequilibrium), marginally, there are far more SNPs associated with gene expression. 

We constructed $\Omega_*^{-1}$ to have a block-diagonal structure and to control the $R^2$. Specifically, we set $\Omega_*^{-1} = D_E \Sigma_E D_E$ where $D_E \in \mathbb{R}^{q \times q}$ is a diagonal matrix with positive entries and $\Sigma_E \in \mathbb{S}_+^{q}$ is a positive definite correlation matrix. 
In our analysis of the GTEx data, we found that on average, the estimated correlation matrix had an approximately $20 \times 20$ correlated block, of which a $10 \times 10$ sub-block was more highly correlated. Thus, we set $$ [\Sigma_{E}]_{j,k} = \left\{ \begin{array}{l l} \rho & \text{ for } (j,k) \text{ where } 11 \leq j \leq 20 \text{, }11 \leq k \leq 20 \text{, and } j \neq k\\
\rho + 0.2 & \text{ for } (j,k) \text{ where } 1 \leq j \leq 10 \text{, } 1 \leq k \leq 10 \text{, and } j \neq k\\
1 & \text{ for } (j,k) \text { where } j = k\\
0 & \text{ otherwise }
\end{array} \right.$$
In the results presented here, we considered $\rho \in \left\{0, 0.1, 0.3, 0.5, 0.7 \right\}$. Given $\Sigma_{E}$ and $\beta_*$, we then generated entries of $D_E$ to determine the $R^2$ for all $q=29$ tissues: we considered $R^2 \in \left\{.01, .05, .10, .20, .40\right\}$. 

Finally, we also randomly assigned missingness to both the training set and validation set responses with missing probability equal to $0.55$, i.e., the missing rate in the GTEx gene expression data we analyzed. For each method, we fit the model to the training data, selected tuning parameters using the validation data, and measured the prediction and variable selection accuracy on the testing data. 

We considered three different methods: two of which can be considered ``complete-case'' estimators. We define the missingness matrix $M \in \mathbb{R}^{n_{\rm train} \times q}$ as: 
$$ M_{i,k} = \left\{ \begin{array}{l l} n_k^{-1/2}  &: y_{i,k}\text{ was observed }\\
  0 & : y_{i,k}\text{ was missing } \end{array} \right. \quad (i,k) \in \left\{1, \dots, n_{\rm train} \right\} \times \left\{1, \dots, q \right\}, $$
 where $n_k$ is the number of subjects with expression observed for the $k$th tissue in the training data. Similarly, let $Y \in \mathbb{R}^{n_{\rm train} \times q}$ be the fully-observed gene expression training data matrix, and $X \in \mathbb{R}^{n_{\rm train} \times p}$ the SNP genotypes for the training data. The methods we compared are:
\begin{itemize}
  \item \texttt{EN}: The tissue-by-tissue elastic net defined as
  $$\argmin_{\beta \in \mathbb{R}^{p \times q}} \left\{  \frac{1}{2n}\|(Y - X\beta) \circ M\|_F^2 + \sum_{k=1}^q \hspace{-1pt}\lambda_k\hspace{-1pt} \left(\alpha_k \sum_{j=1}^p |\beta_{j,k}| + (1-\alpha_k) \sum_{j=1}^p\beta_{j,k}^2\right)\right\},$$
  where for $k = 1, \dots, q$, each $(\lambda_k, \alpha_k) \in \mathbb{R}_+ \times [0,1]$ is chosen to maximize the validation set $R^2$ for the $k$th response variable. This is the default method for estimating eQTL weights in \citet{gamazon2015gene}. 
  \item \texttt{MT}: The exact version of the method proposed by \citet{hu2018statistical}: 
  \begin{equation} \label{hu_estimator}\argmin_{\beta \in \mathbb{R}^{p \times q}} \left\{  \frac{1}{2n}\|(Y - X\beta) \circ M\|_F^2 + \lambda_\beta \mathcal{P}_\beta^{(\alpha)} (\beta) \right\},
  \end{equation}
  where $\lambda_\beta > 0$ and $\alpha \in [0,1]$ are chosen to maximize validation set $R^2$ averaged over all 29 tissues. As explained in the previous section, this is a special case of our method with the restriction that $\Omega = I_q$. 
  \item \texttt{Cov-MT}: The model-based approach we proposed in \eqref{estimator}: 
  $$ \argmin_{\beta \in \mathbb{R}^{p \times q}, \Omega \in \mathbb{S}^q_+} \left\{  \mathcal{F}(\beta, \Omega; Y_O) + \lambda_\beta \mathcal{P}_\beta^{(\alpha)} (\beta) + \lambda_\Omega \mathcal{P}_\Omega(\Omega) \right\},$$
  where tuning parameters $\lambda_\beta > 0, \lambda_\Omega > 0,$ and $\alpha \in [0,1]$ are chosen to maximize the validation set $R^2$ averaged over all 29 tissues. 
\end{itemize}
We also obtained oracle estimators \texttt{Or-EN} and \texttt{Or-MT}, which are equivalent to \texttt{EN} and \texttt{MT} without any missingness. Here \texttt{Or} stands for ``Oracle", i.e., an estimator which has information not available in practice. These estimators replace $M$ with an $\mathbb{R}^{n_{\rm train} \times q}$ matrix with each entry equal to $n_{\rm train}^{-1/2}$. These estimators are meant to compare to the idealized setting where there is no missingness in the response to distinguish between the effects of missingness and the effects of ignoring tissue-tissue correlation. 

Finally, we also considered \texttt{KNN(20)-MT}, which first imputes the missing responses using a weighted mean based on the $20$-nearest neighbors (subjects) and then used the same criterion as \texttt{Or-MT} to estimate $\beta_*$. We also tried imputation via $k$-nearest neighbors with $k \in \left\{2, 5, 10, 50\right\}$, but results did not differ substantially across these choices of $k$, so additional results are omitted. 

We measured performance using three metrics. The metric we used to measure prediction accuracy was average test-set $R^2$. Specifically, for a given estimate of $\beta_*$, we obtain the predicted value of $Y_{\rm test} \in \mathbb{R}^{n_{\rm test} \times q}$, say $\hat{Y} \in \mathbb{R}^{n_{\rm test} \times q}$, and compute
$$ \hat{R}^2 = \frac{1}{q} \left\{\sum_{k=1}^q \left( 1 - \frac{\|Y_{{\rm test},k} - \hat{Y}_{\cdot, k}\|_2^2}{\|Y_{{\rm test}, k} - \bar{Y}_{{\rm train},k}\|_2^2}\right)\right\},$$
where $\bar{Y}_{{\rm train},k} = 1_{n_{\rm test}} \left[ n_k^{-1} \sum_{i=1}^{n_{\rm train}} y_{ik} \mathbf{1}(y_{ik} \text{ was observed})\right]'$ where $1_{n_{\rm test}} \in \mathbb{R}^{n_{\rm test}}$ is a vector of ones and $\mathbf{1}(\cdot)$ denotes the function which equals one if its argument is true and zero if false.  Note that this definition allows for the possibility that the test-set $R^2$ is less than zero which would occur when the training sample mean predicts $Y_{\rm test}$ better than the estimate of $\beta_*$. 
We also measured the linkage disequilibrium (LD)-adjusted true positive rate, which we define as
\begin{equation}
\frac{\sum_{k=1}^q   \left[ \sum_{j=1}^{p} \mathbf{1}\left( [\hat \beta_{*}]_{j,k} \neq 0 \cap [\hat{\beta}]_{l, k} \neq 0 \text{ for any } l \in \left\{l : |{\rm Cor}(X_{\cdot,l},X_{\cdot,j})| > 0.60 \right\} \right) \right]}{\sum_{k=1}^q \sum_{j=1}^p \mathbf{1}(  [\beta_{*}]_{j,k} \neq 0)}.
\end{equation}
The LD-adjustment accounts for the fact that SNP genotypes are often highly correlated. Under our data generating model, we consider an eQTL discovery ``true'' if the selected SNP genotype is moderately correlated with a true eQTL (i.e., predictor whose corresponding regression coefficient is nonzero). 

In addition to true positive rate, we also measured model size, i.e., the proportion of nonzero entries of the estimated regression coefficient matrix. Note that under our data generating models, the true model size is approximately .017, i.e., $(20/p)$, but because SNPs are so highly correlated, much larger estimated models could be expected.

\begin{figure}
\begin{center}
\centerline{\hfill\includegraphics[width=18cm]{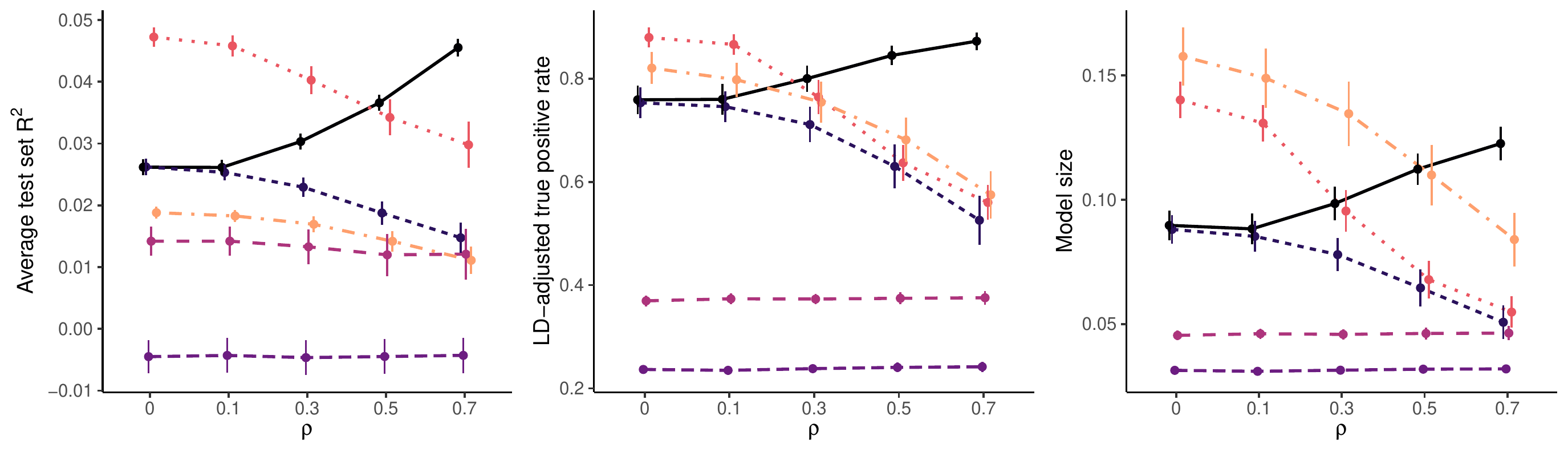}\hfill}
\centerline{\hfill(a)\hfill}
\centerline{\hfill\includegraphics[width=18cm]{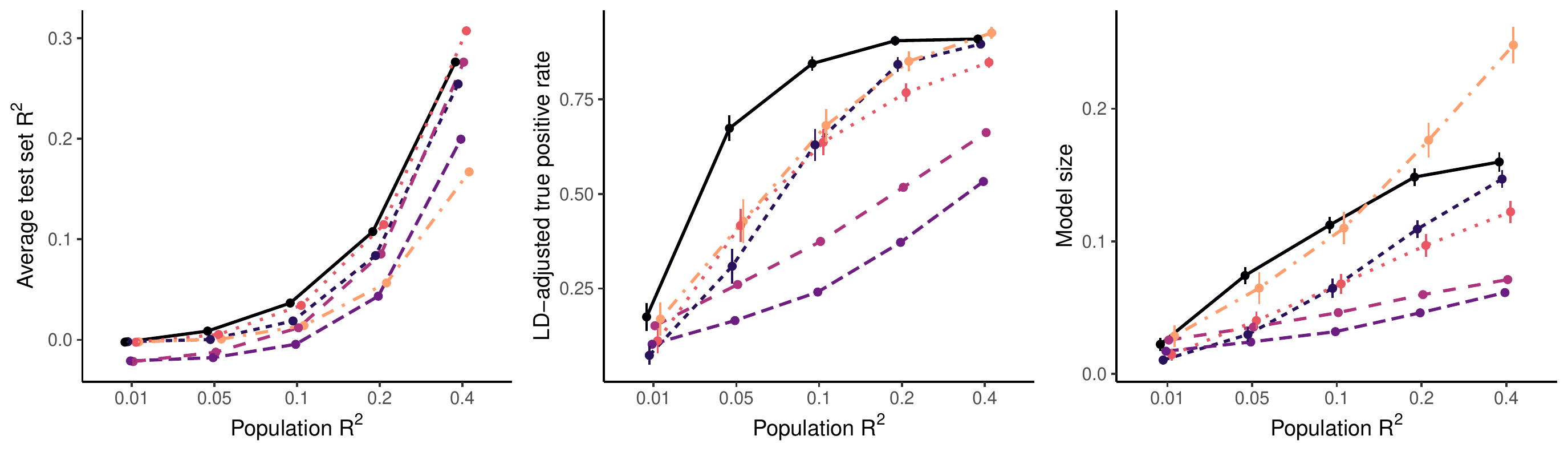}\hfill}
\centerline{\hfill(b)\hfill}
\centerline{\hfill\includegraphics[width=18cm]{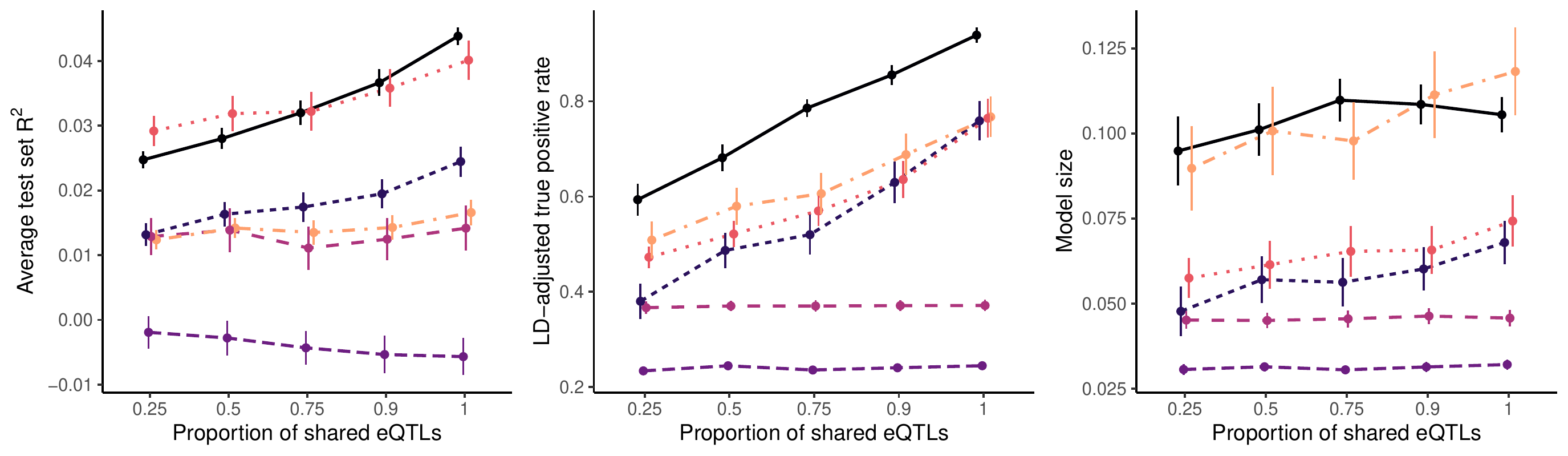}\hfill}
\centerline{\hfill(c)\hfill}
\centerline{\hfill\includegraphics[width=10cm]{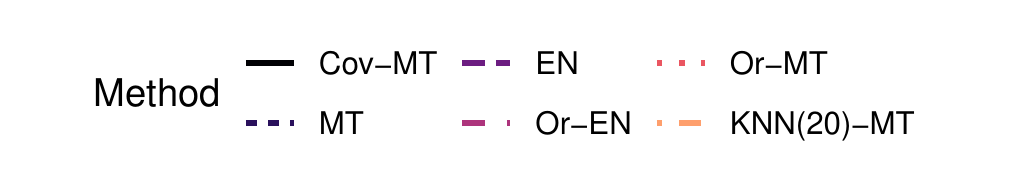}\hfill}
\end{center}
\vspace{-20pt}
\caption{
Average test set $R^2$, average LD-adjusted true positive rate, and average model size (proportion) for six competing methods where: (a) $\rho$, the correlation of the errors varies; (b) the population $R^2$ varies; and (c) the proportion of the twenty eQTLs which are shared across all 29 tissues varies. Error bars denote two times the standard error for each method. Note that the spacing on the horizontal axes is not proportional to the difference in values represented on the horizontal axes. In the leftmost plot of the (b), error bars are not visible due to the range of the vertical axis. Note that throughout, the default settings were $R^2 = 0.1$, the proportional of shared eQTLs was equal to $0.15$, and $\rho = 0.50$.}\label{eq:sim_results}
\end{figure}

\subsection{Simulation study results}
We present complete simulation study results in Figure \ref{eq:sim_results}. In the top row, (a), we present results with tissue-tissue correlation $\rho \in \left\{0.0, 0.1, 0.3, 0.5, 0.7\right\}$ varying, the population $R^2$ fixed at $0.10$, and the proportion of shared eQTLs fixed at $0.75$ (i.e., $s=15$). In this setting, we observe that our method, \texttt{Cov-MT} performs better than all realistic competitors: only \texttt{Or-MT}, the version of \texttt{MT} which does not have any responses missing, outperforms our method when $\rho$ is less than  $0.5$. As one would expect, when expression is nearly uncorrelated ($\rho = 0$), our method \texttt{Cov-MT} performs similarly to the exact version of the method of \citet{hu2018statistical}, which implicitly assumes no tissue-tissue correlation. Remarkably, when $\rho$ is greater than or equal to $0.50$, \texttt{Cov-MT} outperforms even the ``Oracle" methods which have no missingness. In fact, the prediction accuracy of \texttt{Cov-MT} increases as $\rho$ increases, whereas all other methods, which do not explicitly model tissue-tissue correlation, have prediction accuracy remaining constant or slightly decreasing as $\rho$ increases. This demonstrates the benefit of accounting for tissue-tissue correlation in multi-tissue joint eQTL mapping when expression across tissue types can be reasonably assumed to be conditionally dependent. It is also notable that \texttt{EN} performs very poorly: this is partly attributable to the fact that this approach does not leverage potential eQTLs shared across tissues, and thus, has relatively poor variable selection of eQTLs, which is apparent from the results displayed in the middle figure of row (a). 

As $\rho$ increases, the LD-adjusted true positive rate (TPR) of our method tends towards one, whereas for many of the competitors, the true positive rate decreases as $\rho$ increases. This may partly be due to the fact that these methods tend to estimate fewer eQTLs as $\rho$ increases, which is demonstrated in the rightmost figure of row (a).  Notably, all methods tend to yield much larger models than the true model. Finally, we also observe that our method significantly outperforms the 20-nearest neighbor two-step imputation approach (\texttt{KNN(20)-MT}), which first imputes missing values via $k$-nearest neighbor and then fits the \texttt{Or-MT} model to the imputed dataset. This demonstrates the importance of jointly estimating the model parameters and performing expression imputation.
 
In the middle row, (b), of Figure \ref{eq:sim_results}, we present results with $R^2 \in \left\{0.01, 0.05, 0.1,0.2, 0.4\right\}$, the proportion of shared eQTLs fixed at $0.75$, and $\rho = 0.50$ fixed. We observe that our method performs as well or better than  competitors across all settings in terms of average test set $R^2$, except for when $R^2 = 0.40$, in which case \texttt{Or-MT} performs best. Interestingly, our method also performs best in terms of LD-adjusted TPR for small $R^2$, but as $R^2$ approaches $0.40$, many methods tend to perform similarly, though \texttt{KNN(20)-MT} nearly doubled the model size relative to other methods.   It is also notable that even \texttt{EN}, which performed very poorly in the settings displayed in row (a), actually performs better than \texttt{KNN(20)-MT} in terms of prediction accuracy when the population $R^2 = 0.40$.

Finally, in the bottom row, (c), of Figure \ref{eq:sim_results}, results are displayed letting the proportion of shared eQTLs vary with $R^2 = 0.10$ fixed and $\rho = 0.50$ fixed. All methods which can exploit shared eQTLs (i.e., all methods other than  \texttt{EN}, tissue-by-tissue elastic net) improve in prediction accuracy and LD-adjusted TPR as the proportion of shared eQTLs approaches one. Notably, our method performs similarly in terms of prediction accuracy to \texttt{Or-MT}, which is not applicable in practice. In addition, our method, \texttt{Cov-MT}, has higher LD-adjusted TPR than all other methods across all proportions of shared eQTLs. This suggests that accounting for cross-tissue dependence may also improve variable selection accuracy.

\begin{figure}
    \centering
     \centerline{\hfill\includegraphics[width=8cm]{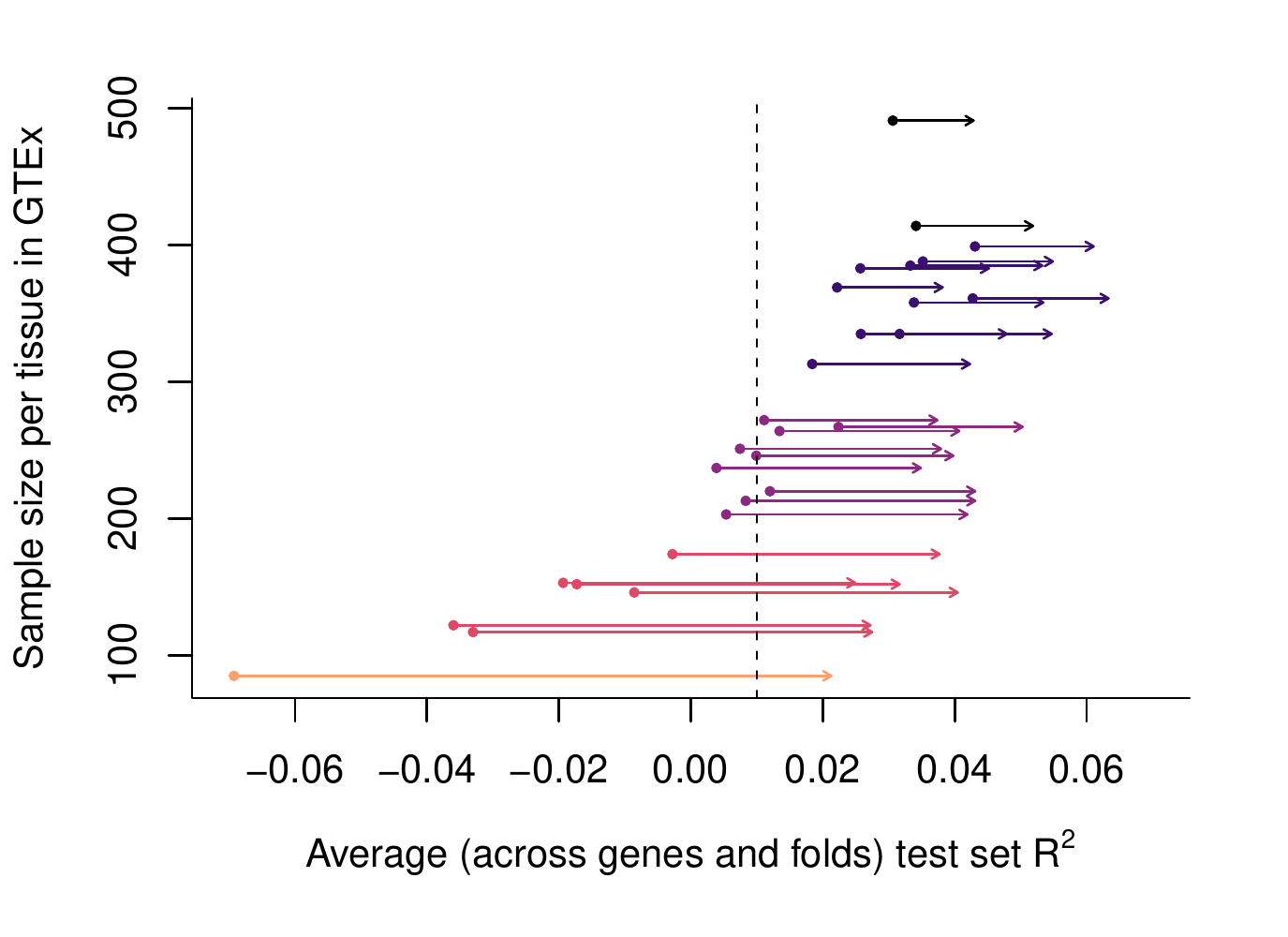}\hfill\includegraphics[width=8cm]{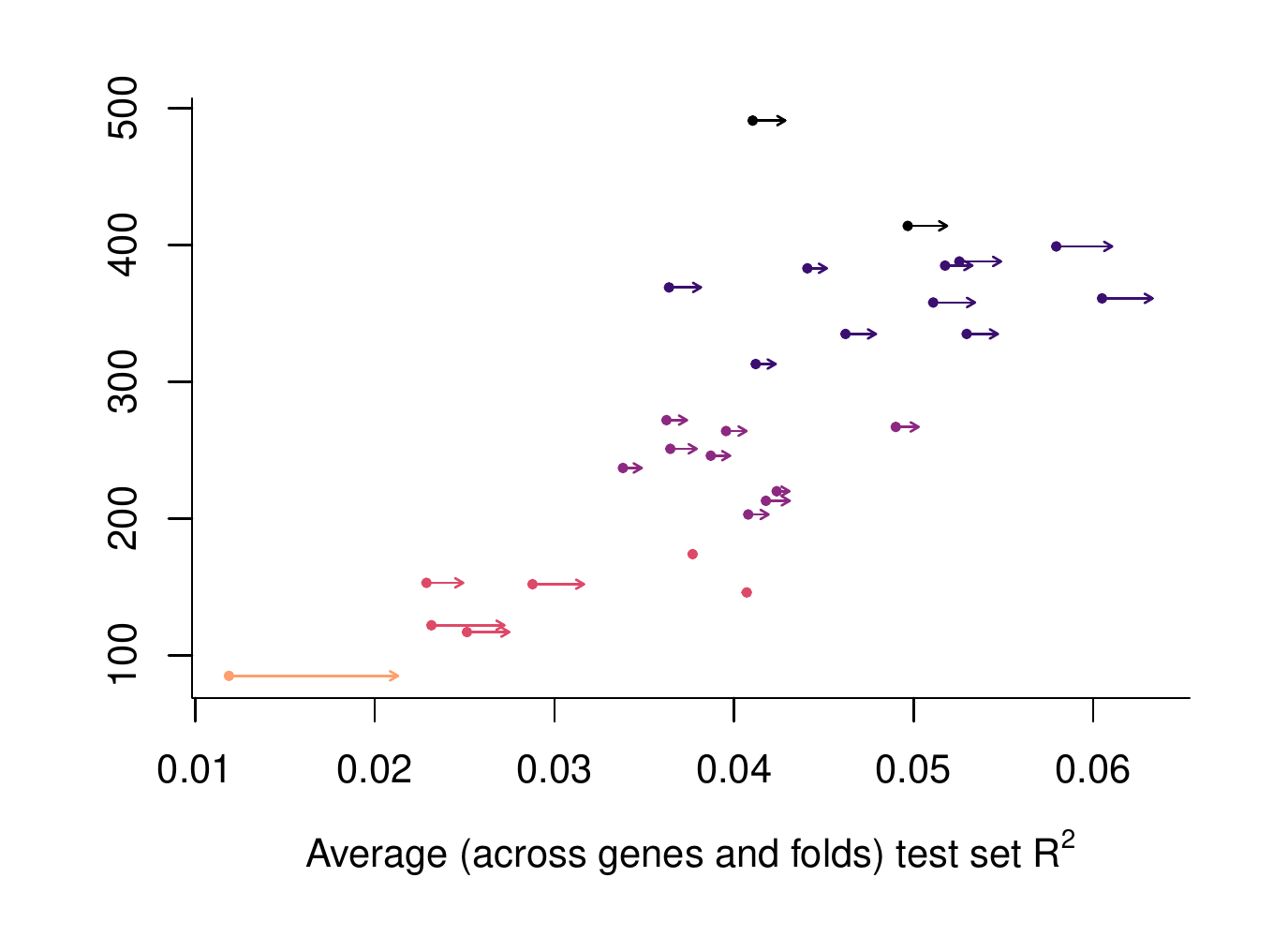}\hfill}
    \centerline{\hfill(a) Comparison to \texttt{EN} \hfill\quad\hfill(b) Comparison to \texttt{MT}\hfill}
     \centerline{\hfill\includegraphics[width=8cm]{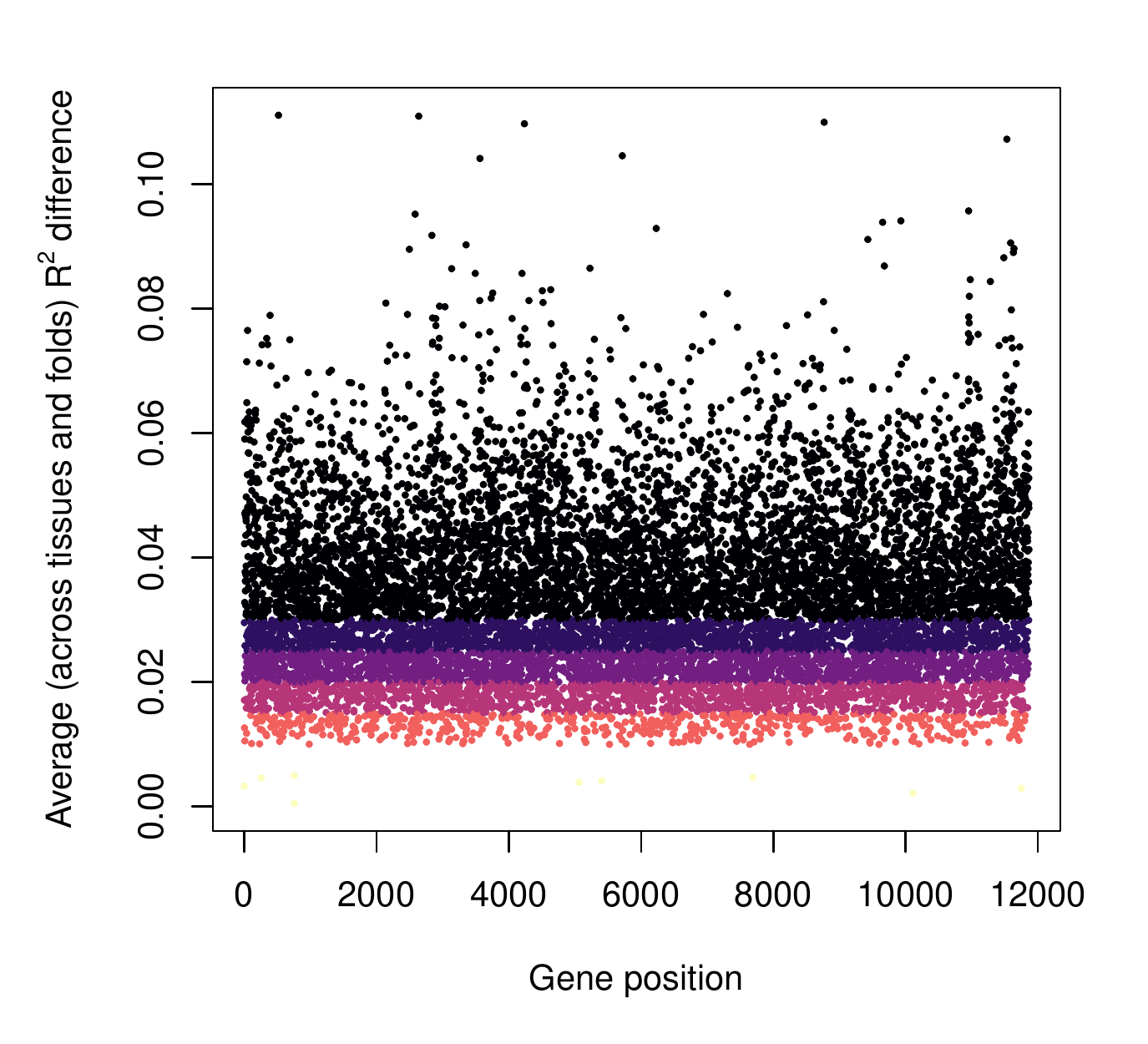}\hfill\includegraphics[width=8cm]{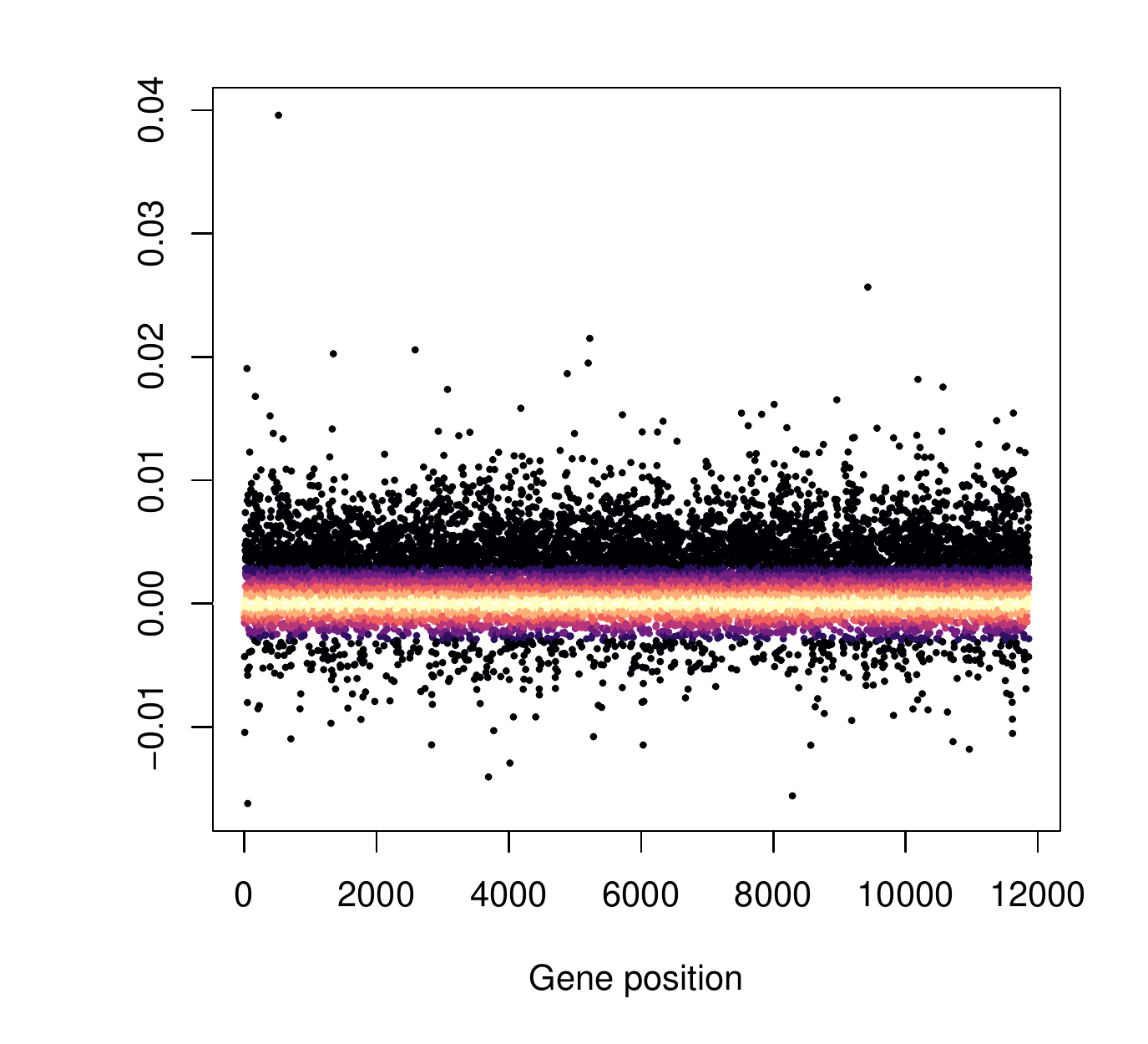}\hfill}
    \centerline{\hfill(c) \texttt{Cov-MT} minus \texttt{EN} \hfill\quad\hfill(d) \texttt{Cov-MT} minus \texttt{MT} \hfill}
    \caption{
    (a, b): Left-out fold $R^2$ averaged across five folds and all genes. In (a), dots correspond to the average $R^2$ for one tissue estimated by \texttt{EN}.  In (b), dots correspond to the average $R^2$ for one tissue estimated using \texttt{MT}. In both (a) and (b), arrow heads point to the average $R^2$ of the same tissue using \texttt{Cov-MT}, and colors correspond to the sample sizes partitioned into intervals of size 100. The points with no visible arrow in (b) had differences less than .0002 in absolute value. (c, d): Differences in left-out fold $R^2$ averaged across five folds and 29 tissues for each gene. In (c), we display the difference using our method minus using \texttt{EN}. In (d), we display the difference using our method minus using \texttt{MT}. In both, color gradients are used to improve visualization -- darker colors are farther from zero. }\label{fig:GTEx_R2}
\end{figure}

\section{Genome-wide multi-tissue joint eQTL mapping in GTEx}
In our analysis of the GTEx data, we focus on 29 types of human tissues (see Figure \ref{fig:missingnessGTEx}). These are tissues which (i) are not sex-specific, (ii) had PEER factors and other covariates available from GTEx, and (iii) are not brain or pituitary gland tissues. Brain and pituitary tissues were omitted because subjects which had expression measured in brain tissue often had no expression measured in the other tissues and vice-versa. 

We obtained RNA-seq data and whole genome-sequencing data from the GTEx Portal (gtexportal.org/home/, v7). We first filtered gene expression separately for each tissue based on read depth, keeping only those genes whose 75th quantile read depth is at least 20. We then kept the intersection of all genes with sufficient read depth across all 29 tissues. Then, we transformed expression using 
$\log_2\left\{(e_{i,j,k} + 1)/q_{0.75}(e_{i,\cdot,k})\right\}$ where $e_{i,j,k}$ is the $j$th gene's read depth for the $k$th tissue for the $i$th subject, and $q_{.75}(e_{i,\cdot,k})$ is the 75th quantile of that subject's read depth within the $k$th tissue across all genes. Finally, we adjusted for age, gender, three genotype principal components, and PEER factors in the same way as described in \citet{hu2018statistical} (i.e., using more PEER factors for tissues with larger sample sizes). Specifically, we regressed the quantile-normalized expression onto these covariates, and used the residuals as our normalized gene expression data for eQTL mapping. 

For each gene, we considered only local eQTLs (cis-SNPs with MAF $\geq .05$), which we defined as SNPs within 500kb of the transcription start or end site of the gene. For each gene, we further prune cis-SNPs until no two SNPs have absolute correlation greater than 0.95.

We performed joint eQTL mapping using the three approaches described in the Simulation studies subsection. Following a similar approach to \citet{hu2018statistical}, we measured prediction accuracy using a five-fold cross validation procedure. That is, each of the five folds was once treated as a testing fold. Of the remaining four folds, three were used to train the model and one was used as a validation set to select tuning parameters. Tuning parameters were selected to maximize the average $R^2$ on the validation set. We repeat this procedure for each of the five folds, with each fold once serving as testing fold and once serving as validation fold.

Results are displayed in Figure \ref{fig:GTEx_R2}, (a) and (b). In Figure \ref{fig:GTEx_R2}(a), our method significantly improves on the tissue-by-tissue elastic net in terms of prediction accuracy, especially for those tissue types with small sample sizes. For example, in the tissue with the smallest sample size (Minor Salivary Gland), the average testing-fold $R^2$ for tissue-by-tissue elastic net was less than $-0.05$, indicating that this approach performed significantly worse than the null model in terms of expression prediction. This suggests that for small sample sizes, the tissue-by-tissue elastic net likely overfits to the training data. Our method, on the other hand, had average testing fold $R^2$ greater than $0.02$ in all tissues, including Minor Salivary Gland. In Figure \ref{fig:GTEx_R2}(b), we also compare \texttt{Cov-MT} to \texttt{MT} and see that incorporating an estimate of the precision matrix $\Omega_*$ improves testing fold $R^2$ averaged over genes and folds by 6.97\% on the tissues we analyzed. Its important to point out that both \texttt{MT} and \texttt{Cov-MT} select tuning parameters to maximize validation set $R^2$ averaged over all tissues, which may partly explain why \texttt{Cov-MT} improves the $R^2$ on tissues with larger sample sizes -- given that these tissues have the highest frequency in the validation folds, they will play the largest role in computing $R^2$ averaged over all tissues. 

\begin{figure}[ht]
    \centering
     \centerline{\hfill\includegraphics[width=15cm]{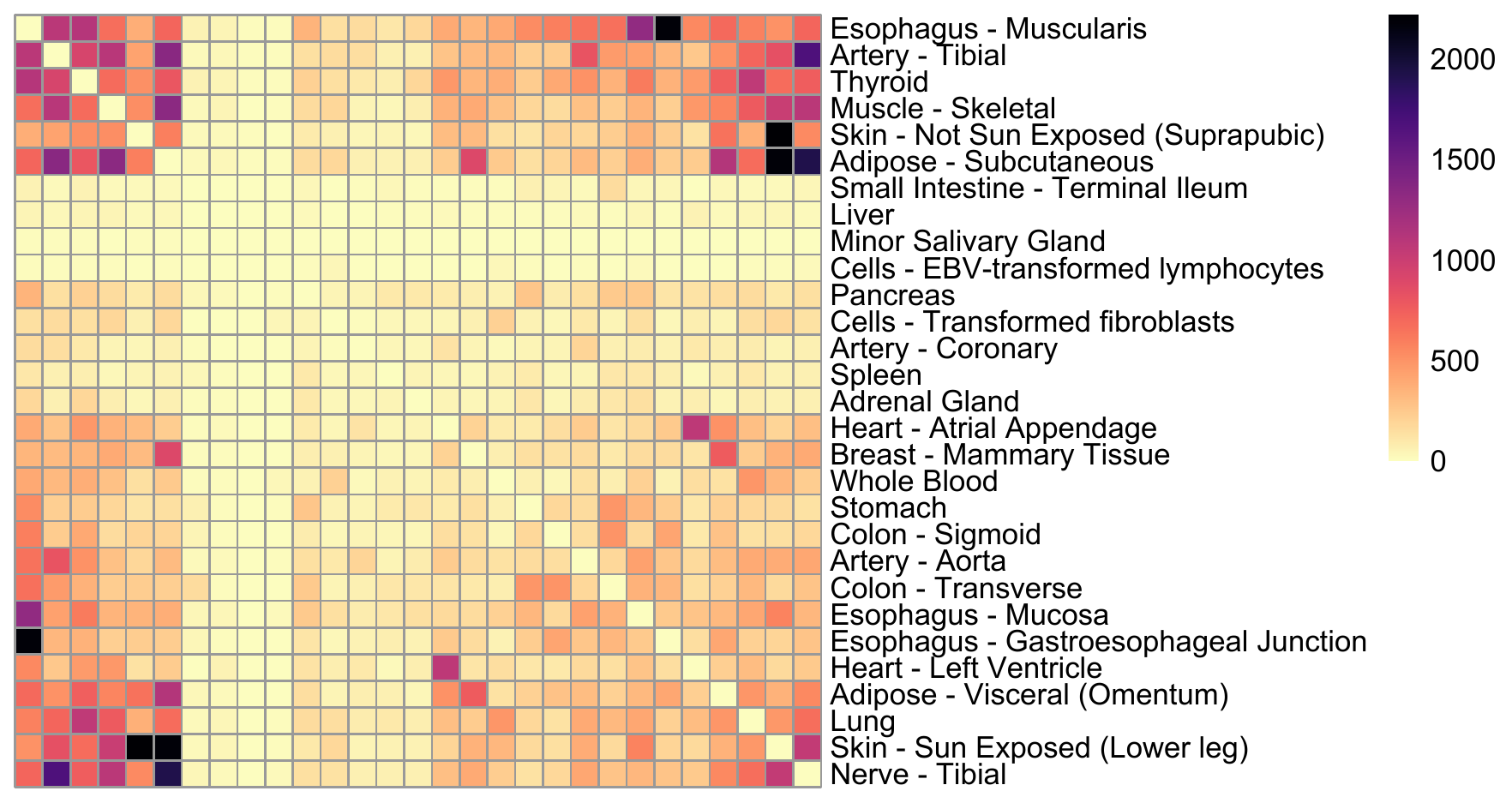}\hfill}
    \caption{
    A heatmap with frequencies of nonzero entries in the estimates of $\Omega_*$. The diagonal has been removed. Tissues were arranged to better visualize pairs which frequently nonzero conditional correlations.}\label{fig:GTEx_Prec}
    \centering
   \end{figure}

In Figure \ref{fig:GTEx_R2}(c) and Figure \ref{fig:GTEx_R2}(d), we display differences in testing-fold $R^2$ averaged across all tissues and folds for each gene we analyzed. For example, each point in Figure \ref{fig:GTEx_R2}(c) denotes the difference in testing-fold $R^2$ averaged over all tissues and folds for our method minus the average using the tissue-by-tissue elastic net. Our method improved on tissue-by-tissue elastic net for nearly every genes; whereas we improved over \texttt{MT} for a majority of genes. In particular, the summary statistics for the difference displayed in Figure \ref{fig:GTEx_R2}(d) (\texttt{Cov-MT} average minus \texttt{MT} average) are: Min = $-0.0161$,  $Q_1 = 3.806 \times 10^{-5}$, $Q_2 = 0.001678$, Mean = $0.001945$, $Q_3 = 0.003685$, and  Max = $0.03959$. 

In Figure \ref{fig:GTEx_Prec}, we display a heatmap of how frequently our method estimated two tissues to be conditionally dependent.  As one would expect, biologically related tissues often had nonzero estimated conditional correlations. For example, the three Esophagus tissues have some of the largest numbers of genes with nonzero tissue-tissue conditional correlations. Similarly, Adipose-Subcantaneous and Skin - Sun Exposed were often conditionally dependent, as were Skin - Not Sun Exposed and Skin - Sun Exposed. Interestingly, many tissues rarely had nonzero estimated conditional correlations with any other tissue: for example, see Small Intestine - Terminal Iluem, Liver, Cells - EBV-transformed lymphocytes, and Minor Salivary Gland.

\begin{figure}[ht]
     \centerline{\hfill\includegraphics[width=15cm]{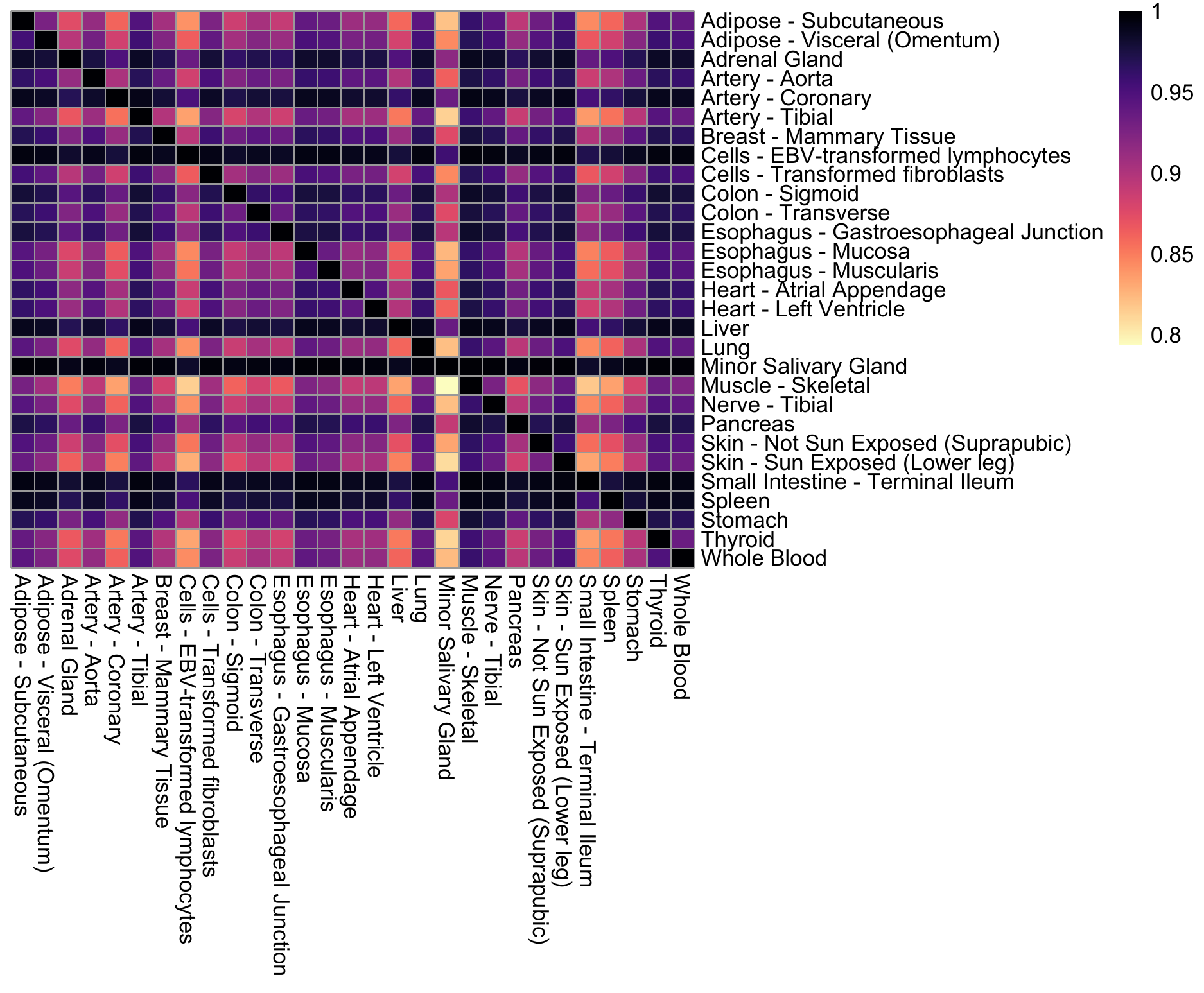}\hfill}
    \caption{
    A heatmap of the proportions of eQTLs shared across tissue types. For a particular row, the intensities correspond to the proportions of total eQTLs for that tissue which are also an eQTL for the column-tissue. For example, the first row indicates that of all Adipose - Subcantaneous eQTLs, approximately 93\% are also eQTLs for Adipose - Visceral and approximately 89\% are also eQTLs for Adrenal Gland.}\label{fig:GTEx_eQTL}
\end{figure}

Another important point of scientific interest is the frequency with which two tissue types share eQTLs. In Figure \ref{fig:GTEx_eQTL}, we display a heatmap displaying the proportion of eQTLs shared between pairs of tissues. The most notable result is that our method (and the method of \citet{hu2018statistical}) tend to estimate that the majority of eQTLs are shared across tissue types. For example, the first row of Figure \ref{fig:GTEx_eQTL} indicates that of all estimated Adipose - Subcantaneous eQTLs, approximately 93\% were also estimated to be eQTLs for Adipose - Visceral and approximately 89\% were also estimated to be eQTLs for Adrenal Gland. Notably,  Minor Salivary Gland and Cells-EBV-transformed lymphocytes were the two tissues which had relatively low proportions of eQTLs shared with each of the other tissues (indicated by the light vertical bands). This may be partly due to the small sample sizes for both of these tissue types, which often led to fewer estimated eQTLs overall.

\section{S-MultiXcan analysis of UKBiobank data}
\subsection{Multi-tissue transciptome-wide association studies}
To demonstrate that our improved multi-tissue joint eQTL mapping method can lead to a higher number of novel gene-level TWAS discoveries, we performed a summary-MultiXcan (S-MultiXcan) analysis following the method proposed in \citet{barbeira2019integrating}. To obtain eQTL weights from the full GTEx dataset, we re-estimated these coefficients using the complete dataset based on the tuning parameters which had the largest average left-out fold $R^2$. We then downloaded UKBiobank GWAS summary statistics from the \citet{Neale} lab for four complex traits: two binary (asthma and heart attack) and two continuous (haemoglobin and platelet count). These traits were selected as we thought no single tissue-type was an obvious candidate for a single-tissue summary-PrediXcan analysis. To compute the LD-matrices needed to perform the S-MultiXcan analysis, we used whole genome-sequencing data obtained from the Genetics and Epidemiology of Colorectal Cancer Consortium (GECCO), which was imputed using the Haplotype Reference Consortium (HRC) reference panel. For each of the three eQTL weight set estimation methods, we tested only those genes which had test set $R^2$ averaged over all folds and tissues greater than zero. The number of genes tested were $10289, 9321,$ $3701$ for \texttt{Cov-MT}, \texttt{MT}, and \texttt{EN} respectively. When testing for association with each phenotype, we adjust for multiple tests using a Bonferonni correction. Thus, \texttt{EN}, which has a smaller number of genes tested, also has a more liberal p-value cutoff.

For each gene, we recorded the S-MultiXcan p-value based on the weights computed using our method, the exact version of the method proposed by \citet{hu2018statistical}, and tissue-by-tissue elastic net (\texttt{Cov-MT}, \texttt{MT}, and \texttt{EN}, respectively). To validate our findings, we also recorded (a) the minimum GWAS p-value amongst the SNPs which had nonzero weights across all methods for each gene, and (b) the minimum GWAS p-value of any SNP with 500kb of the gene transcription start or end site. This way, for any S-MultiXcan discovery, we verified whether this discovery could be attributed to a genome-wide significant eQTL or SNP (based on (a) or (b), respectively) defined as p-value $ < 5 \times 10^{-8}$ or represents a potentially novel finding.  In Table \ref{tab:MultiXcan_Table}, we display the total number of discoveries and the number of novel discoveries for two binary traits and two continuous traits. 

The weight set obtained using \texttt{MT}, the exact version of the method propose by \citet{hu2018statistical}, tended to include a larger number of SNPs. Conversely, our method, which yielded a slightly smaller set of SNPS, has a similar or larger number of significant discoveries than \texttt{MT}. Further, it also has more discoveries which could not be attributed to a GWAS associated eQTL or SNP genotype. For instance, our method identified 18 significant genes, two-thirds did not have an eQTL reaching genome-wide significance, and 3 with no cis-SNPs in the gene reaching the genome-wide significance. In comparison, \texttt{MT} identified 13 genes, among which 7 and 1 gene(s) had no eQTLs or cis-SNPs reaching genome-wide significance, respectively. These genes are listed in Table 2, along with phenotypes with which these genes have been associated in previous GWAS studies (p-value $< 5 \times 10^{-8}$) \citep{buniello2018nhgri}. Notably, most of these genes are associated with phenotyopes related with heart diseases such as coronary artery disease, blood level, cholesterol, high-density lipoproteins(HDL), or low-density lipoproteins (LDL).  When using the more stringent definition of a novel finding based on the p-value of all cis-SNP genotypes, our method still identifies a higher number of novel discoveries for both heart attack and platelet count. In the following section, we further discuss the genes associated with heart attack which could not be attributed to a genome-wide significant eQTL. 

\begin{table}[!h]
\caption{
Number of significant discoveries using different eQTL weights for two binary and two continuous phenotypes. The bottommost two rows (with superscript $*$) denote the number of discoveries which did not have an estimated eQTL reaching genome-wide significance (left) and did not have any cis-SNP reaching genome-wide significance (right). }\label{tab:MultiXcan_Table}
\begin{center}
\makebox[.95\linewidth]{
\scalebox{.95}{
\begin{tabular}{|c||cc|cc|}
  \hline
 & Asthma & Heart Attack & Haemoglobin & Platelet \\ 
  \hline\hline
\texttt{Cov-MT} & 10 & 18 & 890 & 1500 \\ 
  \texttt{MT} & 11 & 13 & 866 & 1478 \\ 
  \texttt{EN} & 6 & 11 & 368 & 600 \\ 
  \hline
  \texttt{Cov-MT}$^*$ & 6/3 & 12/3 & 185/68 & 245/55 \\ 
  \texttt{MT}$^*$ & 3/3 & 7/1 & 183/68 & 228/47 \\ 
  \texttt{EN}$^*$ & 2/1 & 4/1 & 71/36 & 89/37 \\
   \hline
\end{tabular}
}}\end{center}
\bigskip
\caption{
Genes associated with Heart Attack discovered in the S-MultiXcan analysis which could not be attributed to a GWAS associated eQTL. In the rightmost two columns, we provide the phenotypes which with the particular gene has been associated before according to the GWAS Catalog as of October 21st, 2019 \citep{buniello2018nhgri} for all the associations with p-value $< 5 \times 10^{-8}$. The phenotypes for each gene are ranked by the number of times they are associated with this gene (indicated by the numbers in the parenthesis next to the phenotype) and only top two phenotypes are shown for each gene. Bold/underlined genes are those wherein no SNP within 500kb of TSS or TES was genome-wide significant (including those not identified as eQTLs). }
\begin{center}
\makebox[.95\linewidth]{
\scalebox{.85}{
\begin{tabular}{|l|clll|}
\hline
Gene & Chr & Region & Associated phenotype 1 & Associated phenotype 2  \\
\hline
\hline
\textit{AIDA} & 1 & 222668013-222713210 & Coronary artery disease (1) &  
\\
\textit{BROX} & 1 & 222712553-222735196 & & 
\\ 
& & & & \\
\textit{FAM117B} & 2 & 202634969-202769757 & Total cholesterol (2) & LDL (2) 
\\ 
\textit{ICA1L} & 2 & 202773150-202871985 &  Heel bone mineral density (2) & Total cholesterol (1)  
\\
\textit{CARF} & 2 &  202912214-202988263 & Migraine (2) & Coronary artery disease (1) 
\\
\textit{NBEAL1} & 2 & 203014879-203226378 & Coronary artery disease (4)&  Adolescent idiopathic scoliosis (1) 
\\
& & & &\\
\underline{\textit{\textbf{ATG9B}}} & 7 & 151012209-151024499 &  &  
\\
& & & &\\
\underline{\textit{\textbf{LPL}}} & 8 & 19901717-19967259 & Triglycerides (31) & HDL (28) 
\\
& & & &\\
\textit{FURIN} & 15 &  90868588-90883458 & Systolic blood pressure (8) & Schizophrenia (5) 
\\
\textit{FES} & 15 & 90883695-90895776 & Systolic blood pressure (7) & Diastolic blood pressure (4) 
\\
& & & & \\
\underline{\textbf{\textit{TGFB1}}} & 19 & 41301587-41353922 & Blood protein levels (5) & Coronary artery disease (5) 
\\
\textit{BCAM} & 19 & 44809059-44821421& Alzheimer's disease (11) & HDL (5) 
\\
\hline
\end{tabular}
}}
\end{center}
\end{table}

\subsection{S-MultiXcan results}
We have identified multiple genes associated with the occurrence of a heart attack that would be missed in a SNP-by-SNP association analysis. Interestingly, many of these genes are associated with phenotypes related to heart attack in previous GWAS, or are known to have biological functions associated with the occurrence of a heart attack and coronary artery disease. A heart attack occurs when an artery supplying the heart with blood and oxygen becomes blocked. This is closely related to more broadly defined coronary artery disease, which is the narrowing or blockage of the coronary arteries that leads to reduction of the amount of oxygen and nutrients delivered to the heart. Coronary artery disease tends to develop when cholesterol or fatty deposits builds along the artery walls.

Two genes that we identified in Table 2 are not associated with any phenotypes by previous GWAS: BROX and ATG9B. The association with BROX may be due to its close proximity with AIDA. ATG9B, on the other hand, functions in the regulation of autophagy and its expression is induced by hypoxia in endothelial cells \citep{fish2007hypoxia}, and thus, is related to the reduction of oxygen delivery which characterizes coronary artery disease. In fact, endothelial dysfunction is a hallmark of coronary artery disease and another gene identified in our analysis, AIDA, was identified as a coronary artery disease candidate gene by integrative analysis of vascular endothelial cell genomic features \citep{lalonde2019integrative}. Several genes we have identified are involved in different biological processes related with cancer development, implying some connections due to the hypoxia environment shared by narrowed or blocked coronary artery and tumor micro-environment. For example, FES, TGFB and BCAM are all well known cancer related genes involved in signaling for cell growth or apoptosis. CARF regulates cell proliferation and bridges cellular senescence and carcinogenesis \citep{wadhwa2017carf}. A recent report shows that FURIN inhibits apoptosis \citep{zhao2018influence}.

It is important to note that significant associations discovered in this paper do not imply causality \citep{wainberg2019opportunities}. However, as the analysis is concerned with testing of predicted gene expression, it has been shown that such associations are enriched in causal genes and these results should be useful for investigating the disease mechanisms \citep{barbeira2018exploring}. Further, when there is evidence for association with predicted expression of a gene, we may further perform co-localization analysis using methods proposed previously \citep{zhu2016integration, giambartolomei2014bayesian, hormozdiari2016colocalization, wen2017integrating} to examine whether any specific genetic variant is pleiotropic to both the gene expression and disease risk, which may provide evidence of possible causal relationship of the gene and disease.

\section{Discussion}
In this article, we have proposed a new method for obtaining multi-tissue eQTL weights. While our method was motivated by the growing popularity of multi-tissue TWAS using genetically predicted gene expression, it is notable that compared to tissue-by-tissue elastic net, our method yielded eQTL weights which had higher prediction accuracy in every individual tissue we studied. This suggests that even single-tissue PrediXcan analyses could be improved using our estimated weight set. Of course, when to use a multi-tissue test versus a single-tissue test remains an unresolved and important question. Our analyses focused on phenotypes for which no individual tissue seemed an obvious candidate for analyses. 

Another natural application of our method is for imputing unmeasured gene expression, e.g., as was the goal of \citet{wang2016imputing}. Specifically, \citet{wang2016imputing} focused on the case of imputing missing expression in GTEx in individuals with expression measured in a subset of tissues. Our method naturally applies to this problem as the conditional expectation of the missing tissues' expression from \eqref{eq:ConditionalNormal} is easy to compute given estimates of eQTL weights and the cross-tissue error covariance matrix. Furthermore, prediction (ellipsoids) intervals could be constructed using our estimates of the conditional (co)variance of gene expression. Unlike \citet{wang2015joint}, who preselect eQTLs to be used in their prediction model, our approach estimates eQTLs and fits the prediction model jointly. 

Finally, it would be beneficial to extend our methodology to allow for more heavy-tailed error distributions. For example, one could relax the normality assumption in \eqref{eq:Normal_model} and assume that $Y \mid X = x$ follows a multivariate $t$-distribution. One interesting direction along these lines would modify the method of \citet{chen2014regularized} to handle missing data, which is highly nontrivial computationally.

\section*{Supplementary Material}
In the Supplemental Material, we provide the complete algorithms used to solve both \eqref{beta_update} and the exact version of the estimator of \citet{hu2018statistical}, along with discussions of step size parameters. We also include a user-guide for accessing our estimated eQTL weights in order to perform multi-tissue MultiXcan analyses using our weight set. Comprehensive results presented in the article, as well as software to implement the method, are available for download at \href{https://github.com/ajmolstad/MTeQTLResults}{github.com/ajmolstad/MTeQTLResults}.


\section*{Acknowledgments}
The work in this article is funded in part by the National Institutes of Health (CA189532, CA195789, GM105785). The Genotype-Tissue Expression (GTEx) Project was supported by the Common Fund of the Office of the Director of the National Institutes of Health, and by NCI, NHGRI, NHLBI, NIDA, NIMH, and NINDS. The data used for the analyses described in this manuscript were obtained from the GTEx Portal (gtexportal.org/home/, v7) on 04/25/2018.

\renewcommand{\thesection}{}
\bibliography{MTeQTL_arXiv}

\end{document}